\documentclass[preprint, showpacs, floatfix, nofootinbib]{revtex4}
\usepackage{graphicx}
\usepackage{amsmath,amssymb}

\begin{document}

\title{What the collective flow excitation function can tell about the quark-gluon plasma}

\author{Jussi Auvinen}
\email{auvinen@fias.uni-frankfurt.de}
\affiliation{Frankfurt Institute for Advanced Studies (FIAS), Ruth-Moufang-Strasse 1, D-60438 Frankfurt am Main, Germany}

\author{Jan Steinheimer}
\email{steinheimer@fias.uni-frankfurt.de}
\affiliation{Frankfurt Institute for Advanced Studies (FIAS), Ruth-Moufang-Strasse 1, D-60438 Frankfurt am Main, Germany}
\affiliation{Institut f\"ur Theoretische Physik, Goethe Universit\"at, Max-von-Laue-Strasse 1, D-60438 Frankfurt am Main, Germany}

\author{Hannah Petersen}
\email{petersen@fias.uni-frankfurt.de}
\affiliation{Frankfurt Institute for Advanced Studies (FIAS), Ruth-Moufang-Strasse 1, D-60438 Frankfurt am Main, Germany}
\affiliation{Institut f\"ur Theoretische Physik, Goethe Universit\"at, Max-von-Laue-Strasse 1, D-60438 Frankfurt am Main, Germany}

\begin{abstract}
Recent STAR data from the RHIC beam energy scan (BES) show that the midrapidity slope $dv_1/dy$ of the directed flow $v_1$ of net-protons changes sign twice within the collision energy range 7.7 - 39 GeV. To investigate this phenomenon, we study the collision energy dependence of $v_1$ utilizing a Boltzmann + hydrodynamics hybrid model. Calculations with dynamically evolved initial and final state show no qualitative difference between an equation of state with a cross-over and one with a first-order phase transition, in contrast to earlier pure fluid predictions. Furthermore, our analysis of the elliptic flow $v_2$ shows that pre-equilibrium transport dynamics are partially compensating for the diminished elliptic flow production in the hydrodynamical phase at lower energies, which leads to a qualitative agreement with STAR BES results in midcentral collisions. No compensation from transport is found in our model for integrated $v_3$, which decreases from $\approx 0.02$ at $\sqrt{s_{NN}}=27$ GeV to $\approx 0.005$ at $\sqrt{s_{NN}}=7.7$ GeV  in midcentral collisions.
\end{abstract}


\maketitle

\section{Introduction}
In 2010, a beam energy scan program was launched at the Relativistic Heavy Ion Collider (RHIC) to study the features of the QCD phase diagram, in particular the location and the order of the phase transition between the hadronic and QCD matter in the plane of  baryochemical potential $\mu_B$ and temperature $T$. The main observables are the coefficients $v_n$ of the Fourier expansion of the azimuthal angle distribution of final-state particle momenta, which are typically associated with collective flow.

Fluid dynamical calculations have predicted that the slope of the directed flow $v_1$ of baryons will turn negative and then positive again as a function of energy if there is a first order phase transition between hadronic and QCD matter \cite{Rischke:1995pe,Csernai:1999nf,Brachmann:1999xt,Stoecker:2004qu}. 
Qualitatively similar behavior of the midrapidity slope of the net-proton directed flow, $dv_1/dy$, has been found by STAR experiment in the RHIC beam energy scan, with a minimum in the energy interval $\sqrt{s_{NN}}=11.5-19.6$ GeV \cite{Adamczyk:2014ipa}. The elliptic flow $v_2$ is one of the key observables supporting the formation of a strongly coupled quark-gluon plasma (QGP) at the highest energies of RHIC and the Large Hadron Collider (LHC). However, the measured differential elliptic flow $v_2(p_T)$ for charged hadrons remains nearly unchanged from the collision energy range $\sqrt{s_{NN}}=39$ GeV down to 7.7 GeV \cite{Adamczyk:2012ku}, where the formation of hydrodynamically evolving QCD matter is expected to be considerably diminished compared to top RHIC energies. In addition, the preliminary STAR data suggests that the magnitude of the triangular flow $v_3$ remains constant at lower collision energies \cite{Pandit:QM2012}.

We study the collision energy dependence of the collective flow in the RHIC BES range with a hybrid approach, where the non-equilibrium phases at the beginning and in the end of a heavy-ion collision event are described by a transport model, while a hydrodynamic description is used for the intermediate hot and dense stage and the phase transition between the QGP and hadronic matter. This provides a consistent framework for investigating both high-energy heavy ion collisions with negligible net-baryon density and notable quark-gluon plasma phase, and the collisions at smaller energies with finite net-baryon density, where less QGP is expected to form. This approach should thus be ideal for beam energy scan studies.

\section{Hybrid model}
In this study, the transport + hydrodynamics hybrid model by Petersen {\em et al.} \cite{Petersen:2008dd} is utilized. The transport model describing the initial and final state is the Ultrarelativistic Quantum Molecular Dynamics (UrQMD) string / hadron cascade \cite{bass,bleicher}. The transition time to hydrodynamics is defined as the moment when the two colliding nuclei have passed through each other: $t_{\textrm{start}}=2R/\sqrt{\gamma_{CM}^2-1}$, where $R$ represents the nuclear radius and $\gamma_{CM}=1/\sqrt{1-v_{CM}^2}$ is the Lorentz factor.

The (3+1)-D ideal hydrodynamics evolution equations are solved with the SHASTA algorithm \cite{rischke1,rischke2}. The equation of state (EoS) which is utilized most of this study is from Steinheimer {\em et al.} \cite{Steinheimer:2011ea}. It is a combination of a chiral hadronic and a constituent quark model and has the important feature of being applicable also at finite net-baryon densities found at lower collision energies.

The transition from hydro to transport (aka "particlization"), is done when the energy density $\epsilon$ reaches the critical value $\epsilon_C = n\epsilon_0$, where $\epsilon_0=146$ MeV/fm$^3$ is the nuclear ground state energy density. In this study, the values $n=2$ and $n=4$ are used. The particle distributions are generated according to the Cooper-Frye formula from the iso-energy density hypersurface, which is constructed using the Cornelius hypersurface finder \cite{Huovinen:2012is}. The final rescatterings and decays of these particles are then computed in the UrQMD.

\section{Results}

To investigate the sensitivity of the directed flow $v_1(y)=\langle \frac{p_x}{p_T}\rangle_{y_i = y}$ on the order of the phase transition, we run the simulations with a first-order phase transition "Bag model" EoS as an alternative to the above described chiral model EoS which has a cross-over phase transition. The transition from fluid dynamics back to transport happens on an iso-energy density $\epsilon_C = 4\epsilon_0 \approx 0.6 \ \rm{GeV}/\rm{fm}^3$ hypersurface.

\begin{figure}
\centering
\includegraphics[width=5.4cm]{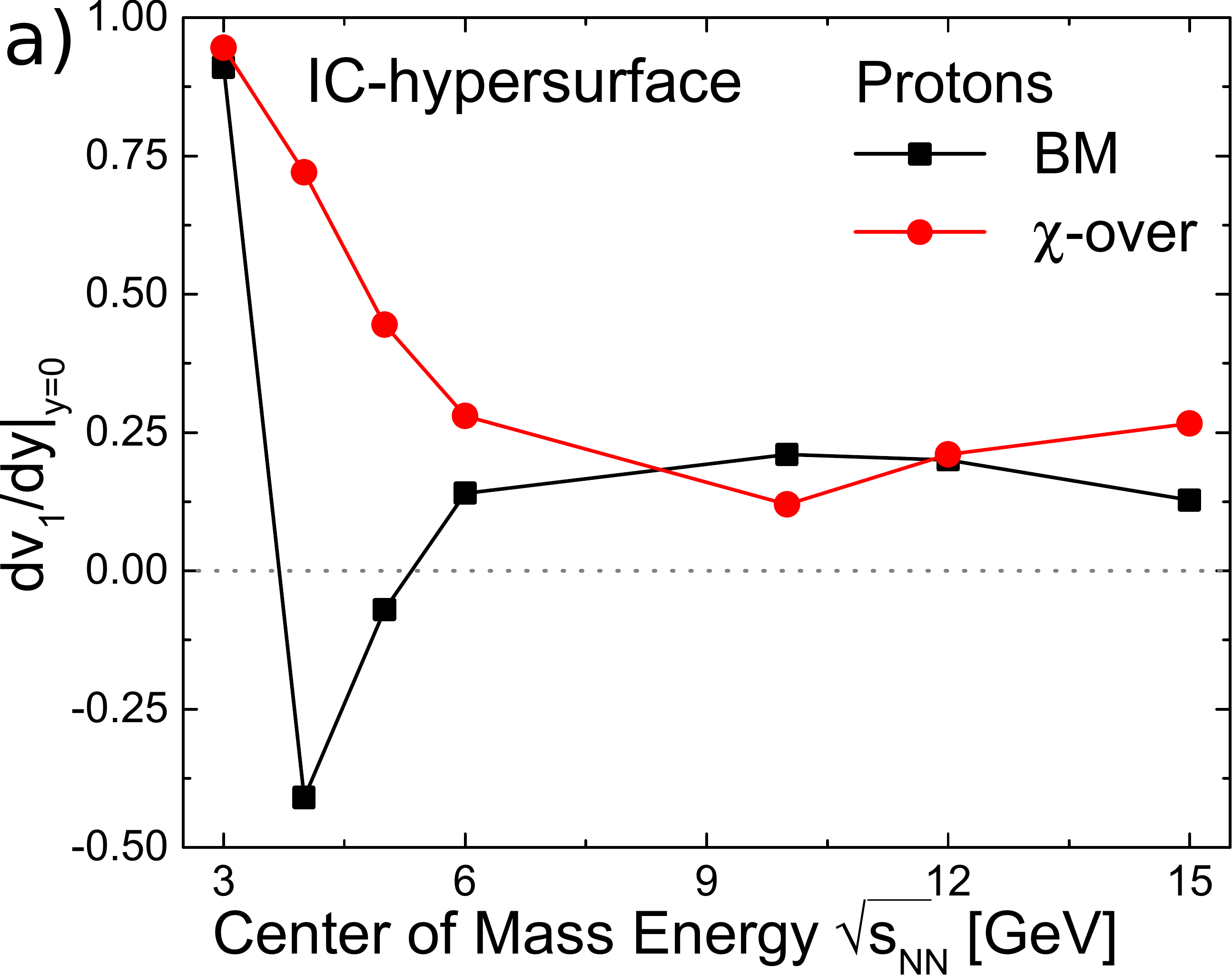}
\includegraphics[width=5.4cm]{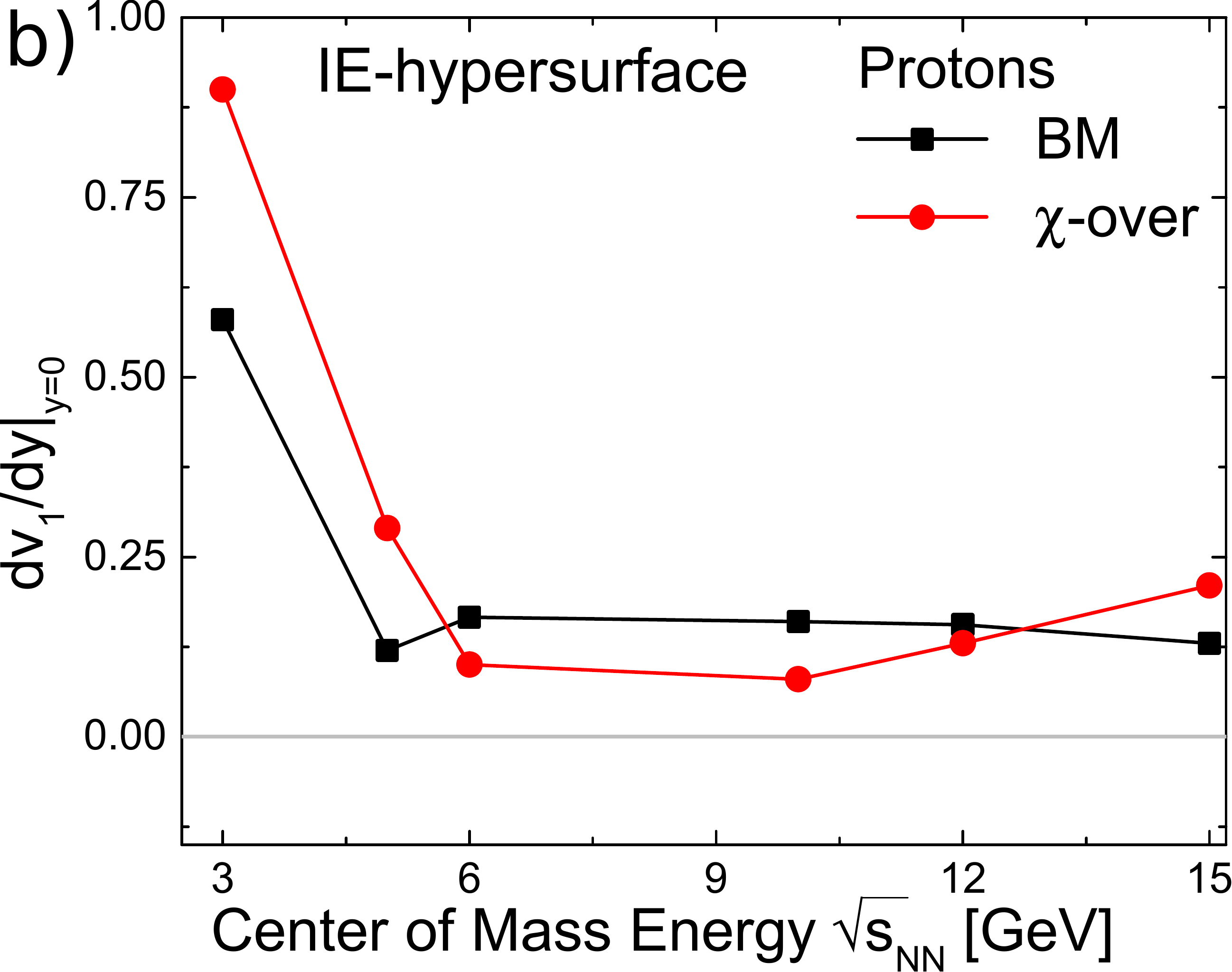}
\includegraphics[width=5.4cm]{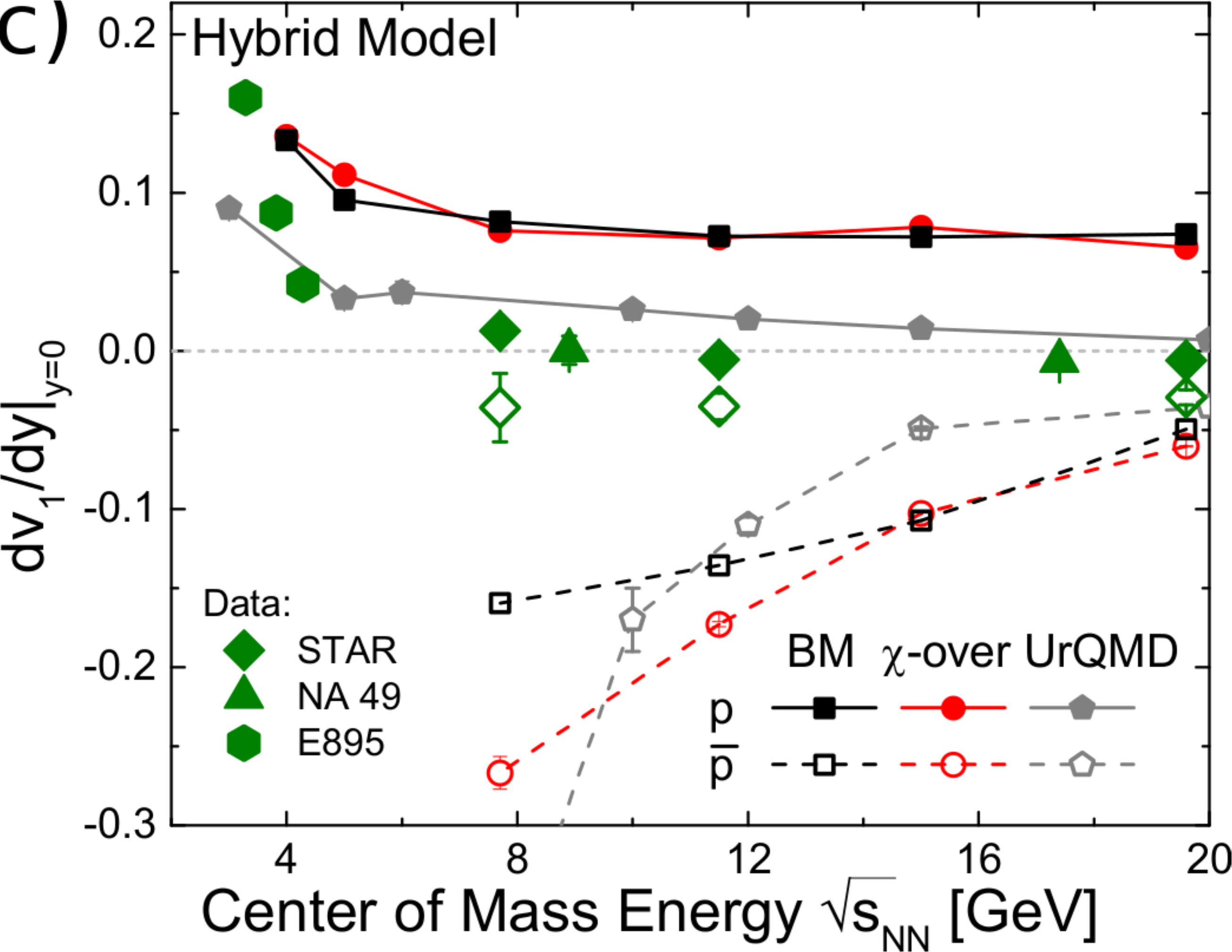}
\caption{a) and b) Slope of $v_1$ of protons around midrapidity $|y|<0.5$ with an equation of state with first-order phase transition (black) and cross-over phase transition (red) for isochronous fluid-to-particles transition hypersurface (a) and iso-energy density transition hypersurface (b) for Au+Au collisions at impact parameter $b$ = 8 fm.  c) Midrapidity slope $dv_1/dy$ of protons (solid symbols) and anti-protons (open symbols) for impact parameter range $b=4.6-9.4$ fm, extracted from the hybrid model calculations with a bag model (black) and crossover EoS (red). Compared with standard UrQMD (grey) and experimental data \cite{Adamczyk:2014ipa,Alt:2003ab,Liu:2000am} (green). Plots from \cite{Steinheimer:2014pfa}.
}\label{Figure_v1}
\end{figure}

To emulate the earlier fluid calculations, we first utilize a ”cold nuclear matter initialization”, where the colliding nuclei are represented by two distributions of energy and baryon density, which respect boosted Woods-Saxon profiles with a central density of saturated nuclear matter $\rho_0 \approx 0.16\,\rm{fm}^{-3}$. The starting point of the simulation is just before the two nuclei first make contact; in the early stage of the collision the kinetic energy of the nuclei is then transformed into large local densities. Figures \ref{Figure_v1}a and \ref{Figure_v1}b show the difference in $dv_1/dy$ between the two equations of state with a cold nuclear matter initialization and the UrQMD afterburner for Au+Au collisions at impact parameter $b$ = 8 fm. The predicted minimum in $dv_1/dy$ as a function of $\sqrt{s_{NN}}$ with a first-order phase transition is clearly observed when using isochronous particlization condition; however, the difference between the two equations of state diminishes greatly when using iso-energy density fluid-to-particles switching condition instead.

Figure \ref{Figure_v1}c shows the result of the full hybrid simulation with the initial non-equilibrium transport phase for the energy dependence of midrapidity slopes of proton and antiproton $v_1$. The directed flow was calculated using events with impact parameter $b=4.6-9.4$ fm, to approximate the $(10-40)\%$ centrality range of the STAR data. The hybrid model overestimates the experimental data and also the pure UrQMD transport result in the whole examined collision energy range. The two EoS are completely indistinguishable in the hybrid simulations, questioning the validity of $v_1$ as a signal of the first-order phase transition. Some possible sources for the difference between the model and the experimental data are the momentum transfer between the spectator particles and the fireball, which is not accounted for in this study, and the method used to determine the event plane, as here $v_1$ was calculated with respect to the reaction plane of the simulation. These uncertainties need further investigation before drawing definite conclusions.

\begin{figure}
\centering
\includegraphics[width=5.5cm]{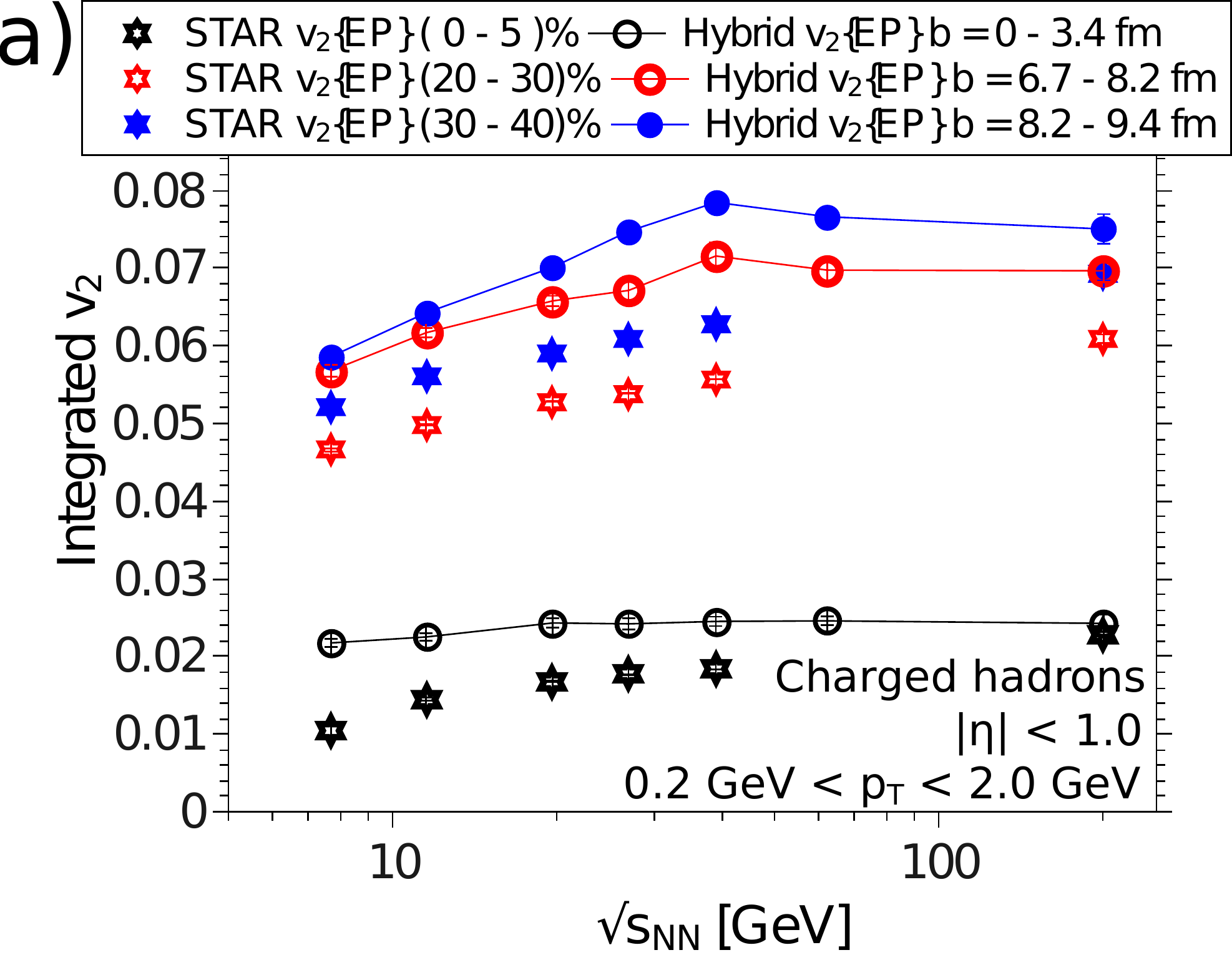}
\includegraphics[width=4.92cm]{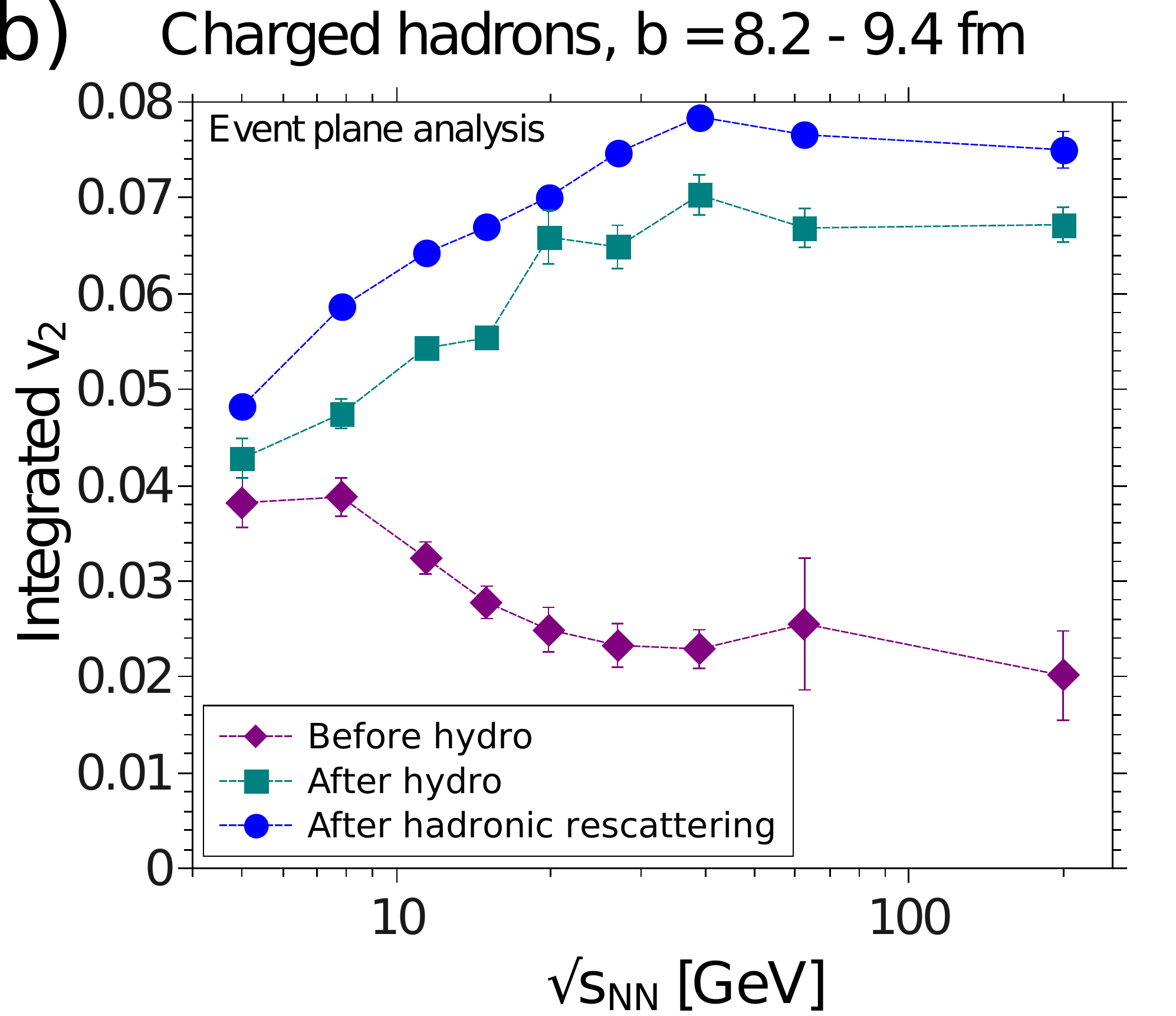}
\includegraphics[width=5.4cm]{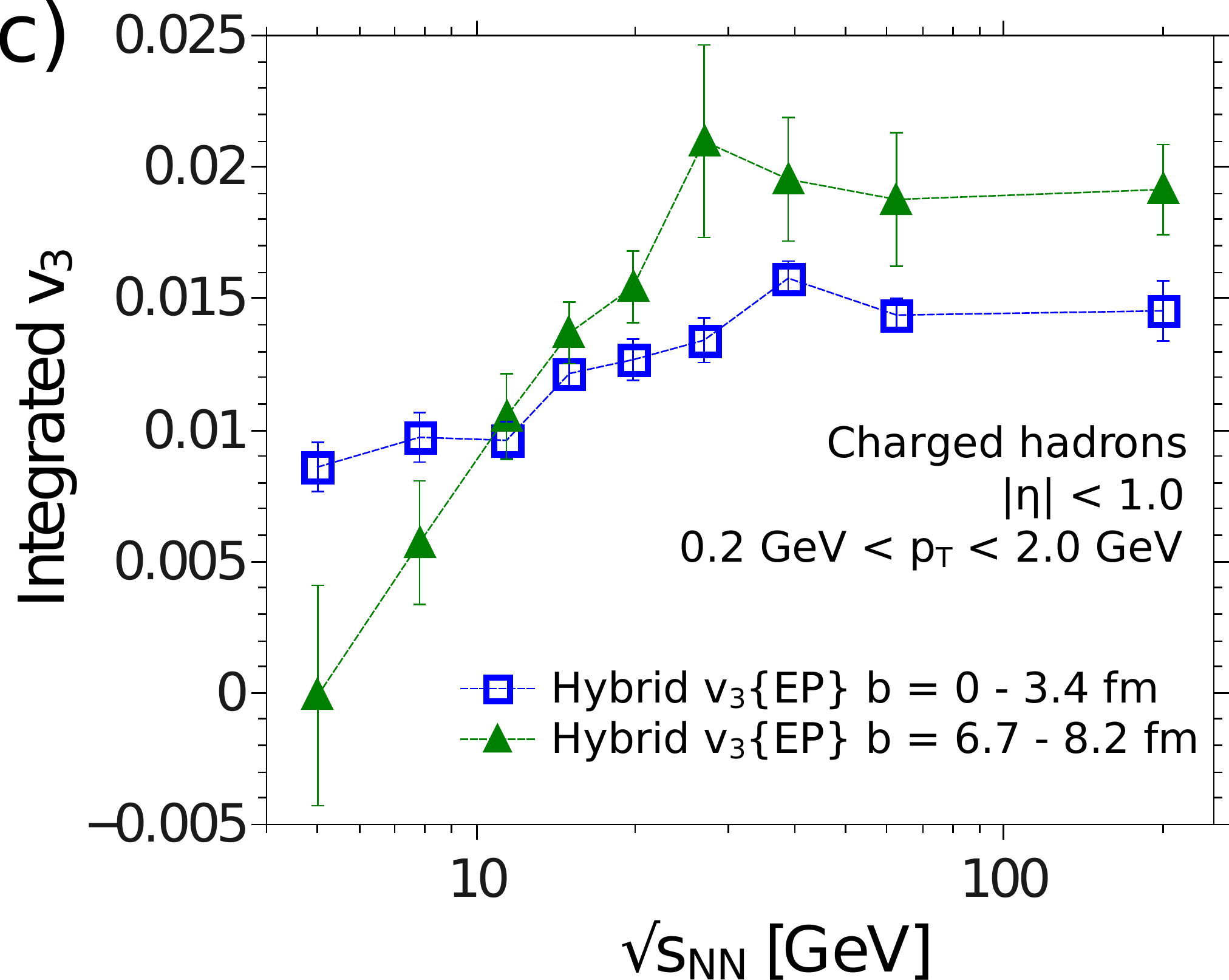}
\vspace{-0.45cm}
\caption{a) Integrated $v_2$ at midrapidity $|\eta|<1.0$ from hybrid model (circles) in collision energy range $\sqrt{s_{NN}}=7.7-200$ GeV at three impact parameter ranges, compared to the STAR data \cite{Adamczyk:2012ku,Adams:2004bi} (stars). b) Integrated $v_2$ for impact parameter range $b = 8.2-9.4$ at $\sqrt{s_{NN}}=5-200$ GeV, at the beginning of hydrodynamical evolution (diamonds), immediately after particlization (squares) and after the full simulation (circles). 
c) Integrated $v_3$ at midrapidity $|\eta|<1.0$ for $b = 0-3.4$ fm and $b = 6.7-8.2$ fm at $\sqrt{s_{NN}}=5 - 200$ GeV. Plots from \cite{Auvinen:2013sba}.}
\label{Figure_v2_v3}
\end{figure}

Utilizing the hybrid model with the crossover EoS, the flow coefficients $v_2$ and $v_3$ are calculated from the particle momentum distributions using the event plane method \cite{poskanzer,ollitrault}. Figure \ref{Figure_v2_v3}a shows the elliptic flow $v_2$ produced in Au+Au -collisions, integrated over the $p_T$ range 0.2 - 2 GeV, compared with the STAR data for the (0-5)\%, (20-30)\% and (30-40)\% centrality classes. In the model these are respectively represented by the impact parameter intervals $b = 0-3.4$ fm, $b = 6.7-8.2$ fm and $b = 8.2-9.4$ fm. Critical energy density $\epsilon_C = 2\epsilon_0$ is used here, as this value has been found to give a reasonable agreement with the experimental data for particle $m_T$ spectra at midrapidity $|y|<0.5$ for energies ranging from $E_{\textrm{lab}}=40$ AGeV to $\sqrt{s_{NN}}=200$ GeV \cite{Auvinen:2013sba,Auvinen:2013qia}. Figure \ref{Figure_v2_v3}b demonstrates the magnitude of $v_2$ at three different times: just before the hydrodynamical evolution, right after the particlization, and the final result after the hadronic rescatterings have been performed in the UrQMD model. 

In the impact parameter range $b = 8.2-9.4$ fm the rescatterings contribute roughly 10\% on the final result. The hydrodynamics produce very little elliptic flow at $\sqrt{s_{NN}} \leq 7.7$ GeV; $v_2$ below $\sqrt{s_{NN}} = 10$ GeV is in practice completely produced by the transport dynamics (resonance formations and decays, string excitations and fragmentation). This initial transport gains importance at lower energies due to the prolonged pre-equilibrium phase. On the other hand, above $\sqrt{s_{NN}}=19.6$ GeV the hydrodynamics are clearly the dominant source of $v_2$.

The simulation results overshoot the experimental data for all collision energies. This suggests that the viscous corrections should be included -- indeed, good results have already been achieved using similar hybrid approach with viscous hydro \cite{Karpenko:2013ama}.  
In the most central collisions below $\sqrt{s_{NN}}=11.5$ GeV, the model deviates from the observed energy dependence; it is likely that the abrupt change from the pre-equilibrium phase to hydro is not a valid assumption in this regime and a more dynamic procedure should be implemented. For the purposes of this study, however, the most important feature is the good qualitative agreement in the midcentral collisions, as here the flow effects are at their largest. 

Based on the above results, it appears that the decrease in the hydrodynamically produced elliptic flow is partially compensated by the increased flow production in the transport phase, and so the observed $v_2$ has weaker collision energy dependence than one would have naively expected. To study this phenomenon further, we do the same analysis for the triangular flow $v_3$, which originates purely from the event-by-event variations in the initial spatial configuration of the colliding nucleons.

As illustrated by Figure~\ref{Figure_v2_v3}c, the $p_T$-integrated $v_3$ increases from $\approx 0.01$ to above 0.015 with increasing collision energy in the most central collisions. However, in midcentrality $b = 6.7-8.2$ fm there is a rapid rise from $\approx 0$ at $\sqrt{s_{NN}} = 5$ GeV to the value of $\approx 0.02$ for $\sqrt{s_{NN}} = 27$ GeV, after which the magnitude remains constant. The energy dependence of $v_3$ in midcentral collisions qualitatively resembles the hydrodynamically produced $v_2$ in Figure~\ref{Figure_v2_v3}b; the viscous medium described by transport smears the anisotropies in the initial energy density profile instead of converting them into momentum anisotropy, and is thus unable to compensate for the lack of ideal fluid described by hydrodynamics. This makes $v_3$ the clearer signal of the presence of low-viscous medium.

\section{Conclusions}

Contrary to the earlier fluid calculations, we have found no difference between an equation of state with a first order phase transition and one with a cross-over phase transition for the midrapidity slope of the directed flow $v_1$, when utilizing the full hybrid model with an iso-energy density switching criterion between hydrodynamics and final transport. Thus $dv_1/dy$ cannot currently be considered as a good signal for the existence of a first-order phase transition. However, there is currently a notable discrepancy between the model results and the experimental data which necessitates further investigation.

We have demonstrated that a hybrid transport + hydrodynamics approach can qualitatively reproduce the experimentally observed behavior of $v_2$ as a function of collision energy $\sqrt{s_{NN}}$. While the $v_2$ production by hydrodynamics is diminished at lower collision energies, this is partially compensated by the pre-equilibrium transport dynamics. Same does not apply to triangular flow $v_3$, which decreases considerably faster, reaching zero in midcentral collisions at $\sqrt{s_{NN}}=5$ GeV. Thus $v_3$ is the better signal for the formation of quark-gluon plasma in heavy ion collisions.

\section{Acknowledgements}

This work was supported by GSI and the Hessian initiative for excellence (LOEWE) through the Helmholtz International Center for FAIR (HIC for FAIR). H.P. and J.A. acknowledge funding by the Helmholtz Association and GSI through the Helmholtz Young Investigator Grant No. VH-NG-822. The computational resources were provided by the LOEWE Frankfurt Center for Scientific Computing (LOEWE-CSC).








\end{document}